# Designing an Algorithm that Detects Fake Amazon Reviews

Seung Ah Choi

## Abstract

Often, there are suspicious Amazon reviews that seem to be excessively positive or have been created through a repeating algorithm. I moved to detect fake reviews on Amazon through semantic analysis in conjunction with meta data such as time, word choice, and the user who posted. I first came up with several instances that may indicate a review isn't genuine and constructed what the algorithm would look like. Then I (with the help from others) coded the algorithm and tested the accuracy of it using statistical analysis and also analyzed it based on the six qualities of code.

## Introduction

### Semantic Evaluation

Here I propose several instances in which certain semantics can point to how the amazon review is not genuine that I will base the code off of.

### *Key For the Greek algorithms

1) 1= the review is fake, 0= the review is not fake.

2) Review 1= R1, Review 2=R2.

3) The abbreviated term for each word is indicated in the box itself

### 1. Look for exaggerated words that may indicate that the review is fake

a) Word Bin

| Too positive(P) | Too negative(N) |
| --- | --- |
| Exceptional | Worst |
| Outstanding | Terrible |
| Astonishing | Appalling |
| Amazing | Disastrous |
| Phenomenal | |

Figure 1

b) Confirmation method: If any of the exaggerated words are included in the review then it may indicate that the review is fake.

c) Algorithm:

    i) Greek: PVN = 1

    ii) English: if the review includes either the too positive term or the too negative term it is fake.

**2. If the review is the exact same as other reviews there is a high chance that the review was copied and pasted multiple times. (In the same way, if there is a way to access the user's previous reviews to see if any are duplicates, that could also be an indicator to figure out what reviews are fake and which are genuine.)**

a) Confirmation method: If there are duplicates of the same review multiple times, it would probably be considered fake.

b) Algorithm:

    i) Greek: -(R1 XOR R2)

    ii) English: if the first review ran through the program is exactly the same as another program then both reviews are fake

**3. Look for professional words that may indicate that the review is fake**

a) Word bin

| Professions (Degree): | Honorifics that may be unnecessary to use for an Amazon review: |
|---|---|
| Ph.D. | Dr. |
| M.D. | Mr. |
| D.D.S. | Mrs. |
| Mention of a Professional Degree including Masters, Post-Masters in fields such as Audiology, Chiropractic, Dentistry, Law, Education, Medicine etc | Captain |
| | Coach |
| | Professor |
| | Reverend |

Figure 2

b) Confirmation method: If the review states to be a profession of higher education that does not necessarily have to mention, the review might be fake because it might be trying to get a free degree.

c) Algorithm

  i) Greek: -(R1 XOR R2)

  ii) English: if the first review ran through the program is exactly the same as another program, then both reviews are fake.

**4. Length of the Review**

a) Confirmation method: If the length of the review is less than two sentences or is a very short sentence/word, the review is probably fake. A good word count could be around more than 10 words and 50 characters in the string.

b) Algorithm

    i) Greek: (Word Count) $\wedge$ (Review Length) = 0

    ii) English: If the review length meets the word count limit, then the review is probably not fake.

## 5. Number of helpful votes

a) Confirmation method: If the number of helpful votes that the review gets is above a certain number, then the review is probably reliable. OR If the number of helpful votes that a certain user gets is significantly less than the number of reviews they gave, that may mean that they are just spamming reviews on Amazon.

b) Algorithm

    i) Greek: (Number of Likes) $\wedge$ (Likes of the Review) = 0

    ii) English: If the review meets the number of likes limit, then the review is probably not fake.

## 6. See if any word from the product title or product category is mentioned in the review

a) Confirmation method: If the number of helpful votes that the review gets is above a certain number, then the review is probably reliable. OR If the number of helpful votes that a certain user gets is significantly less than the number of reviews they gave, that may mean that they are just spamming reviews on Amazon.

b) Algorithm

    iii) Greek: (the product is mentioned) = 1

iv)     English:  If the review contains the name of the product or mentions the product, the review is probably not fake.

## 7. See if the review contains a relevant photo of the product

a)  Confirmation method: If the review includes photos of the product they bought themselves it probably indicates that the review is genuine.

b)  Algorithm

v)      Greek: picture exists = 1

vi)     English:  If the review contains one or more pictures of the product, the review is probably not fake.

<div align="center">

**Overview of the Code**

</div>

Link to code: https://github.com/sachoi613/AmazonReviews/tree/master

Following are specific algorithms that were used in the code:

1.  **Detect presence / absence of a word (or member of word group) & Count unique words:** We used an array list to contain parameters from a method named AmazonReviews. The parameters of the method include the product ID, product title, text, star ratings, helpful votes, and a boolean variable to determine if the user made the purchase of the specific object or not. This array list is also used to go through a tsv file, splitting it up by a tab. Using this the boolean variable to determine if it was a verified purchase which is then set to true. This is used to count the absences of words and the unique words used.

```
 9      public static void main(String[] args) {
10  ArrayList<AmazonReview> allAmazonReviews = parseDataset("data/AmazonReviewDataset.tsv");
11      for (AmazonReview review:allAmazonReviews) {
12          System.out.print(testForFakeReview(review,allAmazonReviews));
13          System.out.println("\t"+review.getStarRating()+"\t" +review.getReviewText());
14  //          System.out.println(review.getSentiment());
15  //          for (int i = 0; i <review.getAllWords().length ; i++) {
16  //              System.out.print(review.getAllWords()[i]+" ");
17  //          }
18  //          System.out.println();
19      }
20
21  }
22      public static ArrayList<AmazonReview> parseDataset(String filename){
23          Scanner scanner;
24          ArrayList<AmazonReview> amazonReviews = new ArrayList<>();
25          try {
26              scanner = new Scanner(new FileInputStream(filename), "UTF-8");
27              scanner.nextLine();
28              while (scanner.hasNextLine()) {
29                  String line = scanner.nextLine();
30                  String [] amazonReview = line.split("\t");
31                  boolean verifiedPurchase = false;
32                  if (amazonReview[11].equals("Y")) verifiedPurchase = true;
33                  amazonReviews.add(new AmazonReview(amazonReview[13],amazonReview[12],amazonRev
34              }
 14
15      public AmazonReview(String reviewText, String reviewHeadline, String productTitle, St
16          this.customerID = customerID;
17          this.reviewDate = reviewDate;
18          this.reviewHeadline = reviewHeadline;
19          this.reviewText = reviewText;
20          this.productTitle = productTitle;
21          this.productCategory = productCategory;
22          this.productID = productID;
23          this.helpfulVotes = helpfulVotes;
24          this.starRating = starRating;
25          this.totalVotes = totalVotes;
26          this.isVerifiedPurchased = isVerifiedPurchased;
27          this.reviewID = reviewID;
28          this.words = getWords();
29          this.wordsSpecificToProduct = getProductSpecificWords();
30          loadSentimentFiles(positiveWords, "data/positive-words.txt");
31          loadSentimentFiles(negativeWords, "data/negative-words.txt");
32      }
33
```

Figure 3

2. **Find most frequently occurring words:** Another skill we used was a txt and tsv file. The txt files were used to store the word bins listed above such as positive and negative words, amazon reviews, and a list of reviews indicating if it is false or not. A Scanner is used to look at each line of the amazon review txt file and trims it. By using the txt files the program is able to search for certain words, resulting in the product.

```
21          }
22      public static ArrayList<AmazonReview> parseDataset(String filename){
23          Scanner scanner;
24          ArrayList<AmazonReview> amazonReviews = new ArrayList<>();
25          try {
26              scanner = new Scanner(new FileInputStream(filename), "UTF-8");
27              scanner.nextLine();
28              while (scanner.hasNextLine()) {
29                  String line = scanner.nextLine();
30                  String [] amazonReview = line.split("\t");
31                  boolean verifiedPurchase = false;
32                  if (amazonReview[11].equals("Y")) verifiedPurchase = true;
33                  amazonReviews.add(new AmazonReview(amazonReview[13],amazonReview[12
34              }
35
36              scanner.close();
37
38          } catch (FileNotFoundException e) {
39              System.out.println("File not found " + filename);
40              return null;
41          }
```

Figure 4

    3. **Count word occurrences:** We also used an if- else statement to determine if the review is negative or positive. It also outputs the predicted rating of the review based on the information. The first if-else statement counts the number of positive and negative words and which also returns if it is extremely positive, negative, or neutral. It uses logical operators (<, >, &&, =) to check if it is higher or lower than a specific percentage. As it uses the logical operators it returns strings suggesting if it is positive or not. After the first if- else statement there is another one leading from it, which uses the specific string which was returned from the previous if- else statement. The if part of the structure uses a string variable called sentiment and checks if that is equal to the specific string returned from the previous if- else statement as it returns the assigned predicted rating for each string. This counts the most frequently used words such as a positive, negative, and neutral category.

```
 84        if (positiveWordCount == 0 && negativeWordCount == 0) return "neutral";
 85        if (positiveWordCount == 0) return "extremely negative";
 86        if (negativeWordCount == 0) return "extremely positive";
 87        double percent = (double) positiveWordCount / (double) negativeWordCount;
 88        if (0.8 < percent && percent < 1.25) {
 89            return "neutral";
 90        } else if (percent > 2) {
 91            return "extremely positive";
 92        } else if (percent < 0.5) {
 93            return "extremely negative";
 94        } else if (percent > 1.25) {
 95            return "positive";
 96        } else if (percent < 0.8) {
 97            return "negative";
 98        }
 99        return "null";
100    }
101
102    public long differenceBetweenSentimentAndStarRating() {
103        String sentiment = getSentiment();
104        long predictedRating = 0;
105        if (sentiment.equals("extremely positive")) {
106            predictedRating = 5;
107        } else if (sentiment.equals("positive")) {
108            predictedRating = 4;
109        } else if (sentiment.equals("neutral")) {
110            predictedRating = 3;
111        } else if (sentiment.equals("negative")) {
112            predictedRating = 2;
113        } else if (sentiment.equals("extremely negative")) {
114            predictedRating = 1;
115        }
```

Figure 5

4. **Find the "distance" between two words in a text:** In the code, a for loop was used to take out the punctuation for more efficiency for the code. It starts with the internet value i which is declared to 0. The logical operator < is used to compare the variable i to the word length. As the last part of the for loop, i is then incremented. In the for loop there is an if statement for if the word contains punctuation (not including specific punctuation and the alphabet). It is then run through more in depth and takes out the punctuation, causing the code to run more easily. This shows the distance between words, slitting them without a punctuation.

```
46      private String stripPunctuation(String word) {
47          String output = "";
48          for (int i = 0; i < word.length(); i++) {
49              if ("abcdefghijklmnopqrstuvwxyz'-".contains(word.substring(i, i + 1))) {
50                  output += word.substring(i, i + 1);
51              }
52          }
53          return output;
54      }
55
```

Figure 6

5. **Count unique words:** To open and close a file, a try- catch block is used to access the file. In the try block, it contains a set of statements where an exception can occur. In the catch block, it handles exceptions and errors that could occur in both blocks. In the code programmed, like in the second paragraph it uses Scanner to look at each line. This reads each line for specific and unique words and categories.

```
55
56      private void loadSentimentFiles(ArrayList<String> wordlist, String filename) {
57          Scanner scanner;
58          try {
59              scanner = new Scanner(new FileInputStream(filename), "UTF-8");
60              for (int i = 0; i < 35; i++) {
61                  scanner.nextLine();
62              }
63              while (scanner.hasNextLine()) {
64                  String line = scanner.nextLine();
65                  wordlist.add(line.trim());
66              }
67              scanner.close();
68          } catch (FileNotFoundException e) {
69              System.out.println("File not found " + filename);
70          }
71      }
72
```

Figure 7

## Test Suit

Following is my thought process for finding a test suit or a Training Set for the algorithm on the internet.

**Things that could potentially be good in a test suite:**

- Name of Product

- Text of the review

- Photos Included in the Review

- Information about the writer of the review

- The writers' previous reviews

- The ratio of total helpful votes the write got versus the number of total reviews posted

- The number of helpful votes on a certain review

- Product number

**Some example Amazon reviews that we found prior to writing the code:**

**Bad Review:**

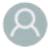

Figure 8

-too short in length

-The content is irrelevant to the product he/she purchased

-There are not that many helpful votes and not a lot of comments

**Profile of a Bad Reviewer:**

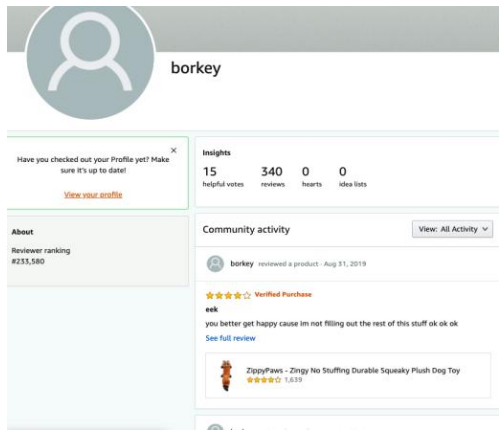

Figure 9

-The ratio of the number of helpful votes to the number of reviews is small

-The profile user name is not credible at all

-All the other reviews are also very short and not legitimate

**Good Review:**

## Customer Review

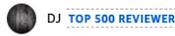

DJ **TOP 500 REVIEWER**

★★★★★ **So Comfy!!**

November 7, 2016

Color: Stone | Size: Standard | **Verified Purchase**

For the last few years I'd been purchasing my reading pillows from Target. They always have them around back to school time. While I've liked them for the most part, over time they would lose their shape and I'd end up sewing them to bring the arms back in. This year I decided to try something new and settled on this one after looking at scores of them on Amazon. I made the decision based on the mid to low price and foam stuffing. I didn't want to invest much more than fourth bucks in a pillow I couldn't see or touch and some of them get up above the hundred dollar range.

When the pillow arrived, I was a bit concerned that it was rolled up and almost completely flat. Fortunately I'd read some reviews that mentioned this and took the advice that I'd need to beat and kneed the pillow pretty vigorously in order to break up the foam clusters. After I finished the pillow had filled out nicely and continued to expand for the next couple days.

Finally, and most importantly, thus pillow is super comfy. It provides plenty of support while remaining pliable, so whenever I need to adjust it conforms to my new position. It also returns to its original shape quickly. It's still early on but I'm really impressed with this pillow. I'll continue to update this review as time goes on if I notice any changes.

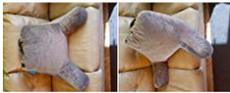

664 people found this helpful

| Helpful |  ⌄ 4 comments | Report abuse | Permalink

Figure 10

-lengthy in text

-picture to support and verify the purchase of the item

-talks about the specific item in the text

-The person is also a top 500 reviewer

-a lot of helpful votes and a decent amount of comments that follows

**Profile of a Good Reviewer:**

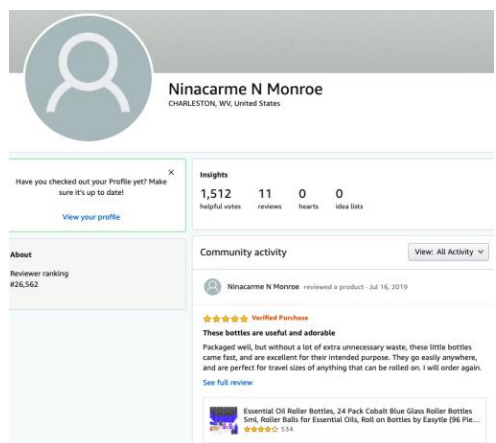

*Figure 11*

-The ratio of helpful votes to the number of reviews is really high

-All the previous reviews are the length and put in great depth of thought

-The name of the profile seems legit, it does not seem like a fake account

**Example of an actual test suite that I found online:**

**Example 1: https://s3.amazonaws.com/amazon-reviews-pds/readme.html**

marketplace    customer_id    review_id    product_id    product_parent    product_title    product_category    star_rating    helpful_votes    total_votes    vine    verified_purchase    review_headline    review_body    review_date

US    18778586    RDIJS7QYB6XNR    B00BY 7X8    122952789    Monopoly Junior Board Game    Toys    5    0    0    N    Y    Five Stars    Excellent!!!    2015-08-31

US    24769659    R36ED1U38IELG8    B00D7JFOPC    952062646    56 Pieces of Wooden Train Track Compatible with All Major Train Brands    Toys    5    0    0    N    Y    Good quality track at excellent price    Great quality wooden track (better than some others we have tried). Perfect match to the various vintages of Thomas track that we already have. There is enough track here to have fun and get creative incorporating your key pieces with track splits, loops and bends.    2015-08-31

US    44331596    R1UE3RPRGCOLD    B002LHA74O 818126353    Super Jumbo Playing Cards by S&S Worldwide    Toys    2    1    1    N    Y    Two Stars    Cards are not as big as pictured.    2015-08-31

US    23310293    R298788GS6I901    B00ARPLCGY    261944918    Barbie Doll and Fashions Barbie Gift Set    Toys    5    0    0    N    Y    my daughter loved it and i liked the price and it came ...    my daughter loved it and i liked the price and it came to me rather than shopping with a ton of people around me. Amazon is the Best way to shop!    2015-08-31

US    38745832    RNX4EXOBBPN5    B00UZOPOFW    717410439    Emazing Lights eLite Flow Glow Sticks - Spinning Light LED Toy    Toys    1    1    1    N    Y    DONT BUY THESE! Do not buy these! They break very fast I spun then for 15 minutes and the end flew off don't waste your money. They are made from cheap plastic and have cracks in them. Buy the poi balls they work a lot better if you only have limited funds.    2015-08-31

US    13394189    R3BPETL222LMIM    B009B7F6CA 873028700    Melissa & Doug Water Wow Coloring Book - Vehicles    Toys    5    0    0    N    Y    Five Stars    Great item. Pictures pop thru and add detail as "painted."  Pictures dry and it can be repainted.    2015-08-31

US    2749569    R3SORMPJZO3F2J    B0101EHRSM    723424342    Big Bang Cosmic Pegasus (Pegasis) Metal 4D High Performance Generic Battling Top BB-105    Toys    3    2    2    N    Y    Three Stars    To keep together, had to use crazy glue.    2015-08-31

US    41137196    R2RDOJQ0WBZCF6 B00407S11Y    383363775    Fun Express Insect Finger Puppets 12ct Toy    Toys    5    0    0    N    Y    Five Stars    I was pleased with the product.    2015-08-31

US     433677         R2B8VBEPB4YEZ7  B00FGPU7U2780517568      Fisher-Price

Octonauts Shellington's On-The-Go Pod Toy         Toys  5     0      0      N      Y

        Five Stars        Children like it         2015-08-31

US     1297934        R1CB783I7B0U52     B0013OY0S0 269360126      Claw Climber

Goliath/ Disney's Gargoyles  Toys  1     0      1      N      Y      Shame on the seller !!!

        Showed up not how it's shown . Was someone's old toy. with paint on it.   2015-08-31

US     52006292       R2D90RQQ3V8LH    B00519PJTW 493486387      100 Foot Multicolor

Pennant Banner         Toys  5     0      0      N      Y      Five Stars      Really liked

these. They were a little larger than I thought, but still fun.   2015-08-31

(...)

**Example 2:**  [http://jmcauley.ucsd.edu/data/amazon/](http://jmcauley.ucsd.edu/data/amazon/)

{"reviewerID": "A2IBPI20UZIR0U", "asin": "1384719342", "reviewerName": "cassandra tu \"Yeah, well, that's just like, u...", "helpful": [0, 0], "reviewText": "Not much to write about here, but it does exactly what it's supposed to. filters out the pop sounds. now my recordings are much more crisp. it is one of the lowest prices pop filters on amazon so might as well buy it, they honestly work the same despite their pricing,", "overall": 5.0, "summary": "good", "unixReviewTime": 1393545600, "reviewTime": "02 28, 2014"}

**Example 3 (The one used for the actual code that includes a ground evidence)**

[https://raw.githubusercontent.com/aayush210789/Deception-Detection-on-Amazon-reviews-dataset/master/amazon_reviews.txt?fbclid=IwAR326w3vt5n51dKP7jKcBT1NQPuEbyehyz_JL8JVbDPwaqKdPYOYrG_5--0](https://raw.githubusercontent.com/aayush210789/Deception-Detection-on-Amazon-reviews-dataset/master/amazon_reviews.txt?fbclid=IwAR326w3vt5n51dKP7jKcBT1NQPuEbyehyz_JL8JVbDPwaqKdPYOYrG_5--0)

DOC_ID      LABEL       RATING       VERIFIED_PURCHASE

        PRODUCT_CATEGORY   PRODUCT_ID       PRODUCT_TITLE

        REVIEW_TITLE      REVIEW_TEXT

**\*Label 1 means fake and Label 2 means true\***

1    __label1__    4    N    PC    B00008NG7N Targus PAUK10U Ultra Mini USB Keypad, Black useful  When least you think so, this product will save the day. Just keep it around just in case you need it for something.

2    __label1__    4    Y    Wireless    B00LH0Y3NM    Note 3 Battery : Stalion Strength Replacement 3200mAh Li-Ion Battery for Samsung Galaxy Note 3 [24-Month Warranty] with NFC Chip + Google Wallet CapableNew era for batteries  Lithium batteries are something new introduced in the market there average developing cost is relatively high but Stallion doesn't compromise on quality and provides us with the best at a low cost.<br />There are so many in built technical assistants that act like a sensor in their particular forté. The battery keeps my phone charged up and it works at every voltage and a high voltage is never risked.

3    __label1__    3    N    Baby    B000I5UZ1Q  Fisher-Price Papasan Cradle Swing, Starlight    doesn't swing very well.    I purchased this swing for my baby. She is 6 months now and has pretty much out grown it. It is very loud and doesn't swing very well. It is beautiful though. I love the colors and it has a lot of settings, but I don't think it was worth the money.

4    __label1__    4    N    Office Products    B003822IRA  Casio MS-80B Standard Function Desktop Calculator    Great computing!    I was looking for an inexpensive desk calcolatur and here it is. It works and does everything I need. Only issue is that it tilts slightly to one side so when I hit any keys it rocks a little bit. Not a big deal.

(...)

10501 __label2__    5    Y    Office Products    B005VCNRA2    SafeT Sleeves RFID Protectors (Total of 8 Sleeves) Fits fine inside a money belt  I purchased this product to separate my credit cards in my money belt for my upcoming trip to Europe.  They fit just fine, and offer a peace of mind from electronic theft anywhere you go. Price was well worth it.

10502 __label2__    5    N    Toys    B00ICAKJJW Power Wheels Nickelodeon Teenage Mutant Ninja Turtles Kawasaki KFX Great fun for little ones. But be sure you fully charge before use.    This is probably one of the most exciting gifts you can give to a young child. Most kids are fascinated with the concept of driving. So having a little vehicle they can safely maneuver is awesome. We purchased this type of vehicle for all 3 of my kids. Everyone of them was thrilled. In fact, we parked them inside the garage facing forward as though they were "parked".<br />This one sits a little high, so the child has to climb up to get on it. But they won't mind.<br />I would highly recommend elbow pads and a helmet. Though we've not seen our 4 wheeler turn over on us yet, but there were a couple of times going over some big bumps on the hilly areas that it looked as though it might tip. So steer your kids away from really bumpy areas and stay on mostly level ground. This is not a TRUE 4 wheeler. It is not designed for rough terrain. It is a TOY.<br />When it first arrives at your house, it might be best not to let the little ones know it's there. It has to be partly assembled and charge overnight (you have to plug it in) before it can be used. If you charge it less than the recommended time, the battery will never hold a charge for long.<br />Being that it runs on an oversize battery, it's only good for about 20 minutes of constant riding before it starts to lose some power.

10503 __label2__ 5 N Toys B0007OTPS2 Ty Beanie Babies Rescue - FDNY Dalmatian Dog Rescuing Rescue I lost my stuffed Rescue toy, and was clearly upset when I did. Then, I had the brilliant idea to look on Amazon. This is the cutest little toy ever and if you get it for any child or dog lover I can guarantee they will love it!

10504 __label2__ 5 Y Home Improvement B008X099V0 Mr. Beams MB982 Wireless Battery Operated Indoor/Outdoor Motion Sensing LED Ceiling Light, White, 2-Pack Battery operated, indoor or outdoor lights. I LOVE these lights! Well, maybe that's too strong a statement but I do like them a whole lot. Simple to install: just insert the batteries (not included), attach the backplate with two screws and attach the light. Sort of like a smoke detector. Motion detector works great, and good lighting.<br /><br />Good customer service if you need to ask a question.

10505 __label2__ 5 Y Tools B0010O748Q SE FS374 All-Weather Emergency 2-IN-1 Fire Starter & Magnesium Fuel Bar (Everything you need to start a fire!) Strike up the band! Great quality and quick shipping I ordered four more of these to go in my survival kits. These are quality made and worth the tiny price! Rick Edwards

10506 __label2__ 5 Y Luggage B005FKVSS8 Rockland Luggage 17 Inch Rolling Backpack, Pink Dot, Medium LOVE THIS ! I'm in nursing school and quickly realized if I didn't get a rolling bag I'd have a bad back before I graduated. I got this because I

LOVE pink. I've gotten a ton of compliments on it and I use it everyday it's still in great shape. I love this bag and will easily use it after school for traveling.

## Corpi Used (Algorithm)

*For the Greek algorithms

1= the review is fake, 0= the review is not fake.

Review 1= R1, Review 2=R2.

The abbreviated term for each word is indicated in the box itself

P = too positive

N = too negative

(Word Count) $\wedge$ (Review Length) = 0

(Number of Likes) $\wedge$ (Likes of the Review) = 0

-(R1 XOR R2)

PVN = 1

I used algorithms related to the different factors that could lead to a review being fake or not.

## Six Qualities of Code

In the code, the method "Amazon Reviews" shows **polymorphism** because various inputs can be put in and the code will still work. The parameters that can be put in this method are string, boolean, long, and integer values/ variables. For example the string values are used to print/show the output of the program being "positive, neutral, and negative". The integer values are associated with the word bins as the percentage of specific negative and positive words there are in the code/review.

```java
public AmazonReview(String reviewText, String reviewHeadline, String productTitle, String productCategory, String productID,
    this.customerID = customerID;
    this.reviewDate = reviewDate;
    this.reviewHeadline = reviewHeadline;
    this.reviewText = reviewText;
```

*Figure 12*

```
D, String customerID, String reviewDate, long helpfulVotes, long starRating, long totalVotes, boolean isV
```

The code shows **completeness** because the code is provable on paper. The code was written out and completed.

The **soundness** in the code/program is logically entailed and provable. As shown beforehand, the code can be written out in algorithms that logically make sense. For example, the variable isVerifiedPurchase is a boolean which can be written out in a logical algorithm to have an output that makes sense.

```java
6   public class AmazonReview {
7       private String reviewText, reviewHeadline, productTitle, productCategory, productID, customerID, reviewDate, reviewID;
8       private long helpfulVotes, starRating, totalVotes;
9       private boolean isVerifiedPurchased;
10      private String[] words;
11      private ArrayList<String> wordsSpecificToProduct;
12      public static ArrayList<String> positiveWords = new ArrayList<>();
13      public static ArrayList<String> negativeWords = new ArrayList<>();
14
```

*Figure 13*

The program is **decidable** because the code will only work when I input some variable. It will not show an infinite loop, it has selection and correct iteration that will give the right output when needed.

```java
46      private String stripPunctuation(String word) {
47          String output = "";
48          for (int i = 0; i < word.length(); i++) {
49              if ("abcdefghijklmnopqrstuvwxyz'-".contains(word.substring(i, i + 1))) {
50                  output += word.substring(i, i + 1);
51              }
52          }
53          return output;
54      }
55
```

*Figure 14*

The program is **correct** because the code does what it should do depending on what we wrote. For example, in the Strings below, and shown in the data set, the public voids returns the variables as they should.

```java
132     public String toString() {
133         return "The review text is " + getReviewText() + "\nThe star rating is " + getStarRating();
134     }
135
136     public String getReviewText() {
137         return reviewText;
138     }
139
140     public String getReviewHeadline() {
141         return reviewHeadline;
142     }
143

144     public String getProductTitle() {
145         return productTitle;
146     }
147
148     public String getProductCategory() {
149         return productCategory;
150     }
151
152     public String getProductID() {
153         return productID;
154     }
155
156     public String getCustomerID() {
157         return customerID;
158     }
159
160     public String getReviewDate() {
161         return reviewDate;
```

*Figure 15*

**Iteration** and **selection** is used for **efficiency** in the code as it shortens the code and runs through it a specific amount of times, allowing there to be more decision making in the code. Loops and if-else statements are used in the code to reduce the amount of time hard coding each specific method and code. This decreases the amount of time spent and helps the programmer include different logical ways to solve the problem. (Specific explanations of the code was elaborated earlier)

```java
102        public long differenceBetweenSentimentAndStarRating() {
103            String sentiment = getSentiment();
104            long predictedRating = 0;
105            if (sentiment.equals("extremely positive")) {
106                predictedRating = 5;
107            } else if (sentiment.equals("positive")) {
108                predictedRating = 4;
109            } else if (sentiment.equals("neutral")) {
110                predictedRating = 3;
111            } else if (sentiment.equals("negative")) {
112                predictedRating = 2;
113            } else if (sentiment.equals("extremely negative")) {
114                predictedRating = 1;
```

*Figure 16*

**Final Algorithm**

 Number of likes = L

Likes of the Review = R

Word Count = W

Review Length = S

Too Positive = P

Too Negative = N

Review 1 = R1

Review 2 = R2.

-(R1 XOR R2) (P V N) (L V R) (W V S)

## Statistical Analysis of Results

**Chi-squared test:**

**n=1999**

|  | **Random model** | **Your results** |
|---|---|---|
| +/+ | **999** | **447** |
| +/- | **999** | **552** |
| -/+ | **1000** | **249** |
| -/- | **1000** | **751** |

*Figure 17. Evaluating test results (using graphs and high-level statistics (either z-test, chi-squared test, or linear regression).*

**Significance level: .05**

The chi-square statistic is 180.2087. The *p*-value is < 0.00001. The result is significant at *p* < .05.

This means that our code is reliable to use and does more than just guessing (when the significance value is α=0.5)

S. Choi, Monta Vista High School, Cupertino, CA 95014